\documentclass[10pt,a4paper]{article}

\usepackage{cite}

\usepackage{times}

\usepackage{epsfig,wrapfig, graphicx,subfigure,multicol}
\usepackage{epstopdf} 

\usepackage{amsmath,amssymb, amsthm}

\usepackage{fancyhdr}

\usepackage[lmargin=2.5cm, rmargin=2.5cm,tmargin=3.50cm,bmargin=3.50cm]{geometry}
\usepackage{indentfirst}

\usepackage{titlesec}
\titleformat{\section}[hang]
  {\centering}{\thesection}{1ex}{\normalsize \textsc}
\titleformat{\subsection}[hang]
  {}{\thesubsection}{1ex}{\normalsize \textit}

\newcommand{\acknowledgement}{\section*{\centering{\textnormal{\normalsize{\textsc{Acknowledgement}}}}}}

\renewcommand{\thesection}{ \normalsize \textnormal{\Roman{section}.}}
\renewcommand{\thesubsection}{\normalsize \textnormal{\textsc{\textit{\Alph{subsection}.}}}}

\def\e{\begin{equation}}
\def\f{\end{equation}}
\def\_#1{{\bf #1}}

\def\.{\cdot}

\begin{document}

\title{\large \textbf{Chessboard Mushroom-Type Metasurface for Beamforming Applications}}
%
\def\affil#1{\begin{itemize} \item[] #1 \end{itemize}}
\author{\normalsize \bfseries A. Abraray$^{1,2}$, R. Pereira$^{1}$, K. Kaboutari$^{2}$ and  S. Maslovski$^{1,2}$}
\date{}
\maketitle
\thispagestyle{fancy} 
\vspace{-6ex}
\affil{
  \begin{center}\normalsize $^1$Instituto de Telecomunicações, 3810-193, Aveiro, Portugal\\
  	$^2$University of Aveiro, Dept. of Electronics, Telecommunications and Informatics, Campus
    Universitário de Santiago, 3810-193, Aveiro, Portugal\\
    abdelghafour.abraray@ua.pt
  \end{center}}

\begin{abstract}
\noindent \normalsize
\textbf{\textit{Abstract} \ \ -- \ \
 A reconfigurable microwave reflectarray metasurface (MS) is investigated for beamforming applications. The reflected beam direction is changed by applying external dc voltages, which create a reflection phase gradient on the structure. The studied MS comprises a chessboard-like array of metallic patches placed over a grounded dielectric slab with metallic vias connecting the patches to the controlling lines. Tunability is achieved with nonlinear capacitive loads (varactors) inserted between the corners of the metallic patches. The MS is studied analytically, numerically and experimentally, from which reshaping of the radiation pattern is observed according to the applied control voltages on the MS elements. It is shown that the proposed MS-based reflectarray with just 3-by-10 elements is already sufficient to redirect the beam in different directions.
}
\end{abstract}

\section{Introduction}
Progress in wireless communications brings new challenges and use cases imposed by the growing requirements of the developing 5G and 6G mobile networks. As is well known, propagating microwaves suffer from disturbances and attenuation incurred by the multipath fading, wave diffraction, material penetration loss, atmospheric absorption, etc. To mitigate such unwanted effects, reconfigurable elements can be introduced into the propagation channel. Such elements are smart beamforming antennas and other systems based on the intelligent reflecting metasurfaces (MS). In the recent years, the concept of the smart MS has emerged as a revolutionary technology that may allow for radiation pattern engineering and beam steering while enabling a fully versatile control over the propagation and scattering of the electromagnetic (EM) waves. Programmable metasurfaces (PMS), also known as reconfigurable intelligent surfaces (RIS), can be used to generate practically arbitrary radiation patterns making the propagation environments adjustable at will.

For example, the authors of~\cite{Bao} proposed a new type of the MS with both phase and amplitude modulations that can simultaneously control multiple beams using adjustable amplitude and phase responses of the MS unit cells. In \cite{Miao}, a method of using exact incident phase for the design of reflective beamforming MS is presented. The PMS has the ability to control the radiation magnitude in both depth (for beam focusing) and the orientation of the radiated wavefront (beam steering), which results in a three dimensional (3D) beam control capability with such an MS. The radiation pattern flexibility realized with the help of reflecting MS can be achieved by using different techniques such as inserting varactor diodes, PIN diodes, microelectromechanical switches (MEMS) or by incorporating active material layers into the structure, like graphene, liquid crystals (LC), vanadium dioxide (VO$_2$), or germanium antimony telluride (GST). In this work, we achieve the radiation pattern flexibility in a wide frequency range by incorporating varactor diodes into a chessboard-patterned Sievenpiper mushroom-type MS. This relatively simple technique results in a low-cost, compact and low-power solution that enables a continuous control of the phase response of the PMS unit cells.

\section{Beamforming with the reflecting PMS}
Beamforming is an advanced technology that directs the signal of a wireless transmitter towards a specific receiver. Rather than having the signal spread in all directions as with broadcasting transmitting antennas, beamforming antennas aim to realize optimal connection channels, which are faster, have higher signal quality, and are more reliable than without any beamforming method. Beamforming technics can also offer an improved solution to reduce the signal interference levels, extend coverage (especially, for highly mobile applications), reduce the energy consumption and improve the systems' capacity.

Realizations of the MS based on the Sievenpiper's mushrooms typically employ a capacitive patch grid placed on top of a thin metal-backed dielectric substrate. Such structures operate as reflecting MS. On the other hand, when operation in the transmission regime is needed, one may use stacks of varactor-loaded patch grids optimized for minimal reflectance and maximum transmittance. Both reflecting and transmitting MS are capable of generating variations in the phase of passing EM waves. The varactor-loaded MS can be adaptively controlled by means of low-power electronic circuits in order to redirect the impinging microwaves towards specified directions or focus them at specified locations.
\begin{figure}
	\centering
    \includegraphics[width=0.4\textwidth]{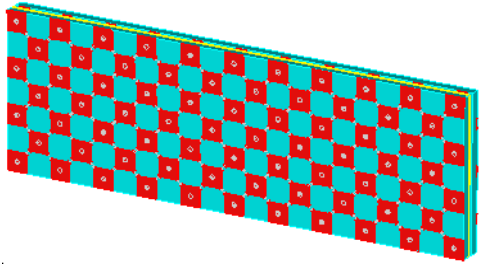}
	\caption{Sievenpiper mushrooms-based chessboard-patterned MS as realized in SIMULIA CST Studio Suite.
	}
	\label{fig:CST_design}
\end{figure}
In what follows, we consider a reflecting MS such as the one shown in Fig.~\ref{fig:CST_design}. When a plane or spherical wave is incident on such a reflecting MS formed by $M\times N$ unit cells, each of the unit cells will scatter the impinging wave with certain amplitude and phase. The local reflection coefficients at the MS unit cells are complex functions of the EM interactions between the cells and the distribution of the incident EM field. The angular dependence of the scattered electric field magnitude produced by such MS in the far zone can be written as (assuming that the MS lies in the $xy$-plane):
\begin{equation}
    \label{E(u,v)}
    E(u, v) \propto \sqrt{1-u^2/(k_0d)^2} \left|\sum_{n=0}^{N-1}\sum_{m=0}^{M-1}\frac{|\Gamma_{mn}|e^{i\phi_{mn}} e^{i(m - (M - 1)/2)u + i(n - (N - 1)/2)v-i k_0 R_{mn}}}{R_{mn}}\right|,
\end{equation}
where $k_0 = \omega/c$ and $\Gamma_{mn} = |\Gamma_{mn}|e^{i\phi_{mn}}$ is the complex reflection coefficient at the $mn^{\rm th}$ element of the MS that relates the scattered field at that point to the incident field. In this formula, we also consider the phase and amplitude variation of the incident field on the MS, as for an impinging spherical wave. The variables $u$ and $v$ are defined as $u = k_0 d \sin\theta\cos\varphi$ and $v = k_0 d \sin\theta \sin\varphi$, where $\theta$ and $\varphi$ are the angles in the spherical coordinate system, whose $z$-axis is normal to the MS, and $d$ is the period of the MS, which is the same along the $x$ and $y$ axes.

In Eq.~(\ref{E(u,v)}), $R_{mn} = \sqrt{R_f^2+[(m -(M - 1)/2)^2+(n - (N - 1)/2)^2]d^2}$ is the radial  distance from the source antenna to the $mn^{\rm th}$ element of the MS. $R_f$ is the distance from the source antenna to the MS. Because the far-field radiation pattern (expressed in the $(u,v)$ coordinates) of a rectangular aperture is equivalent to the Fourier transform of the aperture wave field, the necessary amplitude and phase distributions of the scattered field on the MS can be found with the inverse Fourier transform. This can be done for an arbitrary function $F(u,v)$ that represents a desired radiation pattern. In this way the necessary local scattering phases can be expressed as follows:
\begin{equation}
    \phi_{mn} = \angle\sum_{l=0}^{N-1}\sum_{k=0}^{M-1}C_{kl} \, e^{-2 \pi i\left (\frac{(m-(M-1)/2) (k- (M-1)/2)}{M}+\frac{(n-(N-1)/2) (l-(N-1)/2)}{N}\right)} + k_0 R_{mn},
    \label{phi_mn}
\end{equation}
where $C_{kl} = F\left(\frac{2\pi (k-(M-1)/2)}{M}, \frac{2\pi (l-(N-1)/2)}{N}\right)$ are samples of the desired complex radiation pattern with the indices $k = 0,\dots, M-1$ and $l = 0,\dots, N-1$. In a realistic programmable MS the phases of the local reflection coefficients, $\phi_{mn}$, can be set up to shape the radiated beam in a specific direction accordingly to Eq.~(\ref{phi_mn}). This is why such structures have been named Programmable Metasurfaces (PMS). Hovewer, because the local scattering magnitudes $|\Gamma_{mn}|$ obtained with the Fourier transform method may exceed the $[0;1]$ interval, only the phase information obtained by this method can be used reliably in beamforming with passive PMS. 

\subsection*{Simulations and Numerical Verification of the Theory}
To validate the theoretical beamforming approach outlined above, in our SIMULIA CST Studio Suite simulations we use an array of 30 elements in the $3\times 10$ configuration (Fig.~\ref{fig:CST_design}). The period of the structure is $6.3$~mm and the structure is designed to operate at around $5$~GHz. The array is illuminated by a plane wave, and the electric field far-field pattern is obtained by the full-wave EM simulations. The modulation of the local reflection phase is achieved through the change of the effective surface impedance and the resonant frequency of the MS unit cells. In the real MS, applying appropriate bias voltages to the unit cells loaded by varactors produces a reflection phase gradient along the MS surface, thus altering the direction of the reflected wave front. The necessary phases are obtained by changing the state of each unit cell. Therefore, by engineering the bias voltage distribution along the MS the desired wave front control can be achieved. 

\begin{figure}
	\centering
	\subfigure[]{\label{fig:rad_pattern1_sim}
	    \includegraphics[width=0.35\textwidth]{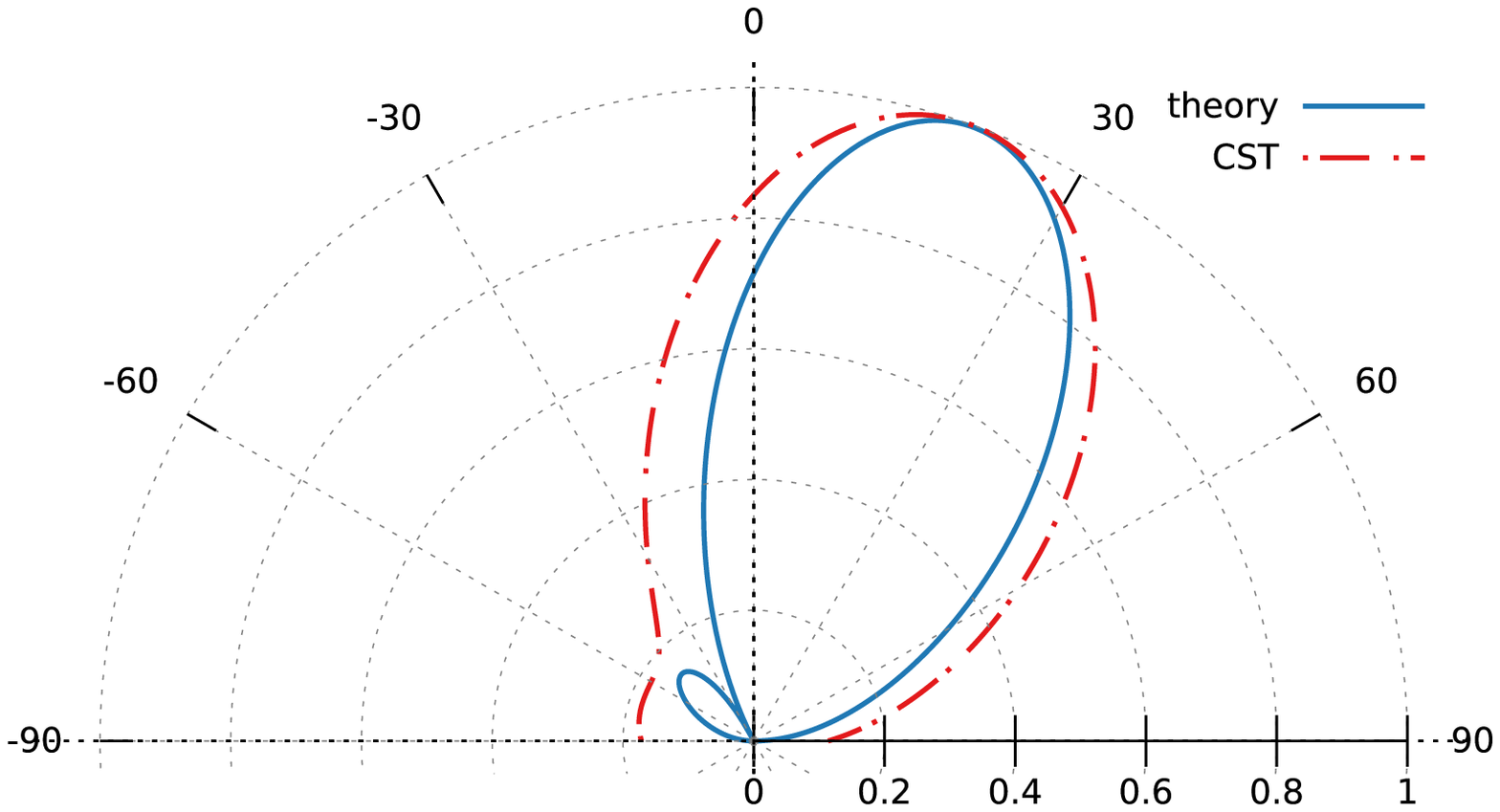}
	}
	\qquad
	\subfigure[]{\label{fig:rad_pattern2_sim}
	    \includegraphics[width=0.35\textwidth]{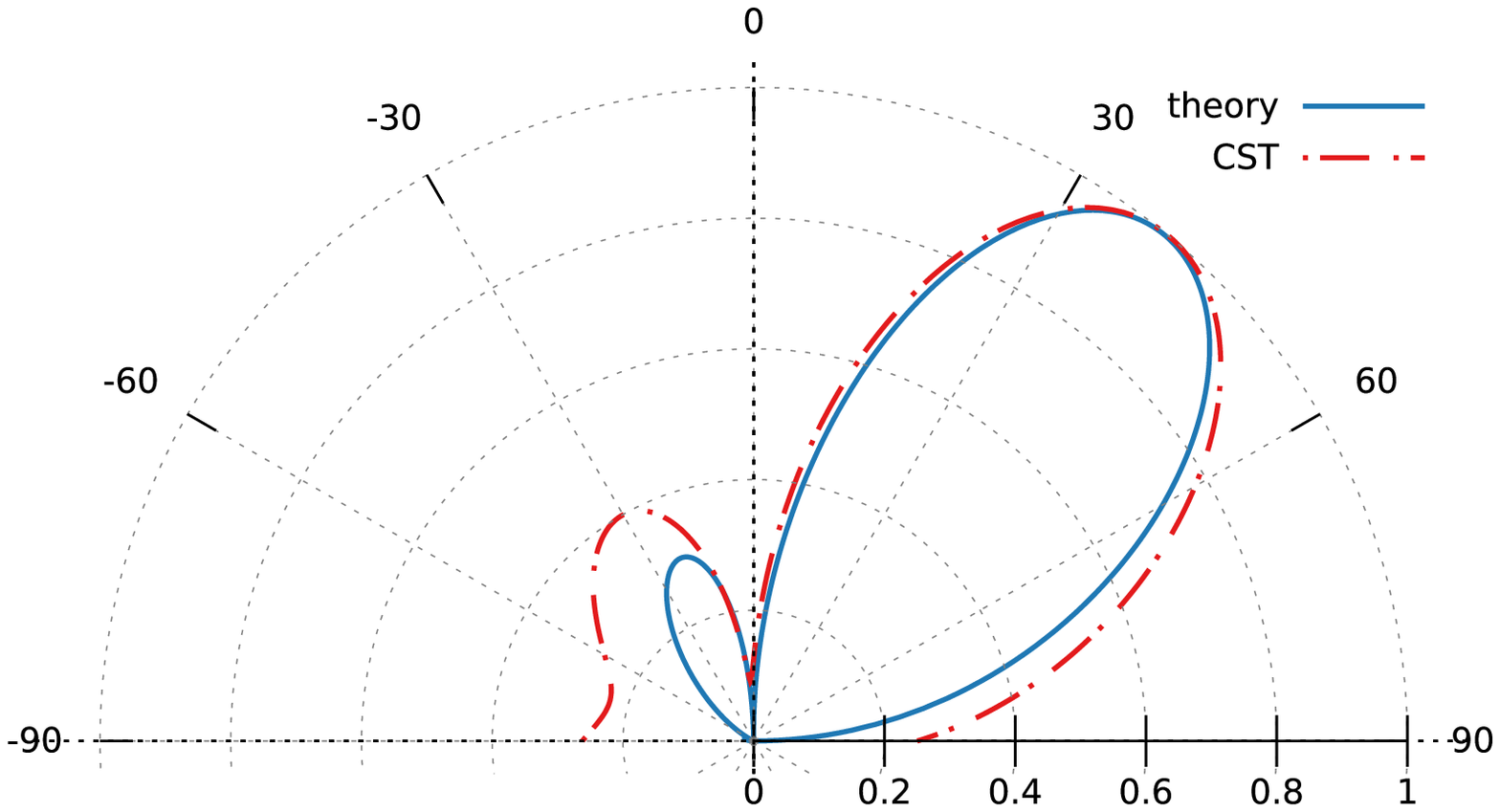}
	}
	\vspace{-3mm}\\[-1mm]
	\caption{Radiation patterns in the $xz$-plane  produced by the PMS formed by and array of 3-by-10 unit cells.}
	\label{fig:rad_pattern_sim}
\end{figure}

We have modeled different phase distributions along the MS that correspond to realistic bias voltage gradients. An example of the realized far-field pattern in the $xz$-plane ($\theta = 0$ correspond to the normal to the MS) is shown in Fig.~\ref{fig:rad_pattern_sim} for the phase variation from $-50^\circ$ to $50^\circ$ (varactor capacitance from $0.25$ to $0.2$~pF) [Fig.~\ref{fig:rad_pattern1_sim}], where the main beam points at the direction of $(\theta_0,\varphi_0) = (19^\circ,0^\circ)$, and for the phase gradient of $-150^\circ$ to $60^\circ$ (0.31 to 0.15~pF) [Fig.~\ref{fig:rad_pattern2_sim}], where the main beam points at $(\theta_0,\phi_0) = (39^\circ,0^\circ)$. In addition to the main beams, parasitic sidelobes occur when the main beam is pointed at $|\theta_0| > 30^\circ$. We observe that the theoretical and simulated radiation patterns are in a good agreement. 

\subsection*{Experimental Verification}
\begin{figure}[b]
	\centering
	\subfigure[]{\label{fig:PCB1}\includegraphics[width=0.3\textwidth]{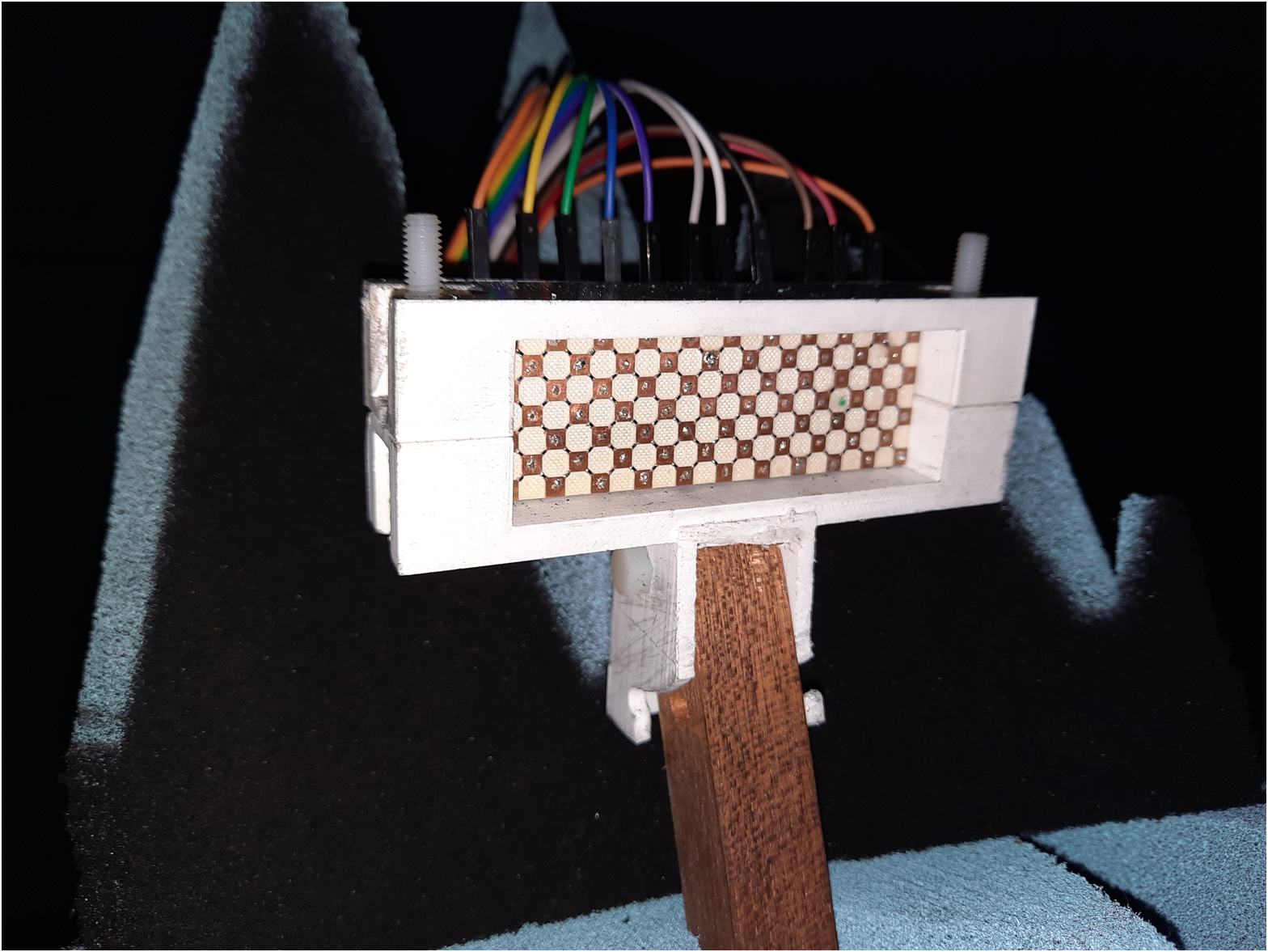}}
	\qquad \subfigure[]{\label{fig:setup1}\includegraphics[width=0.3\textwidth]{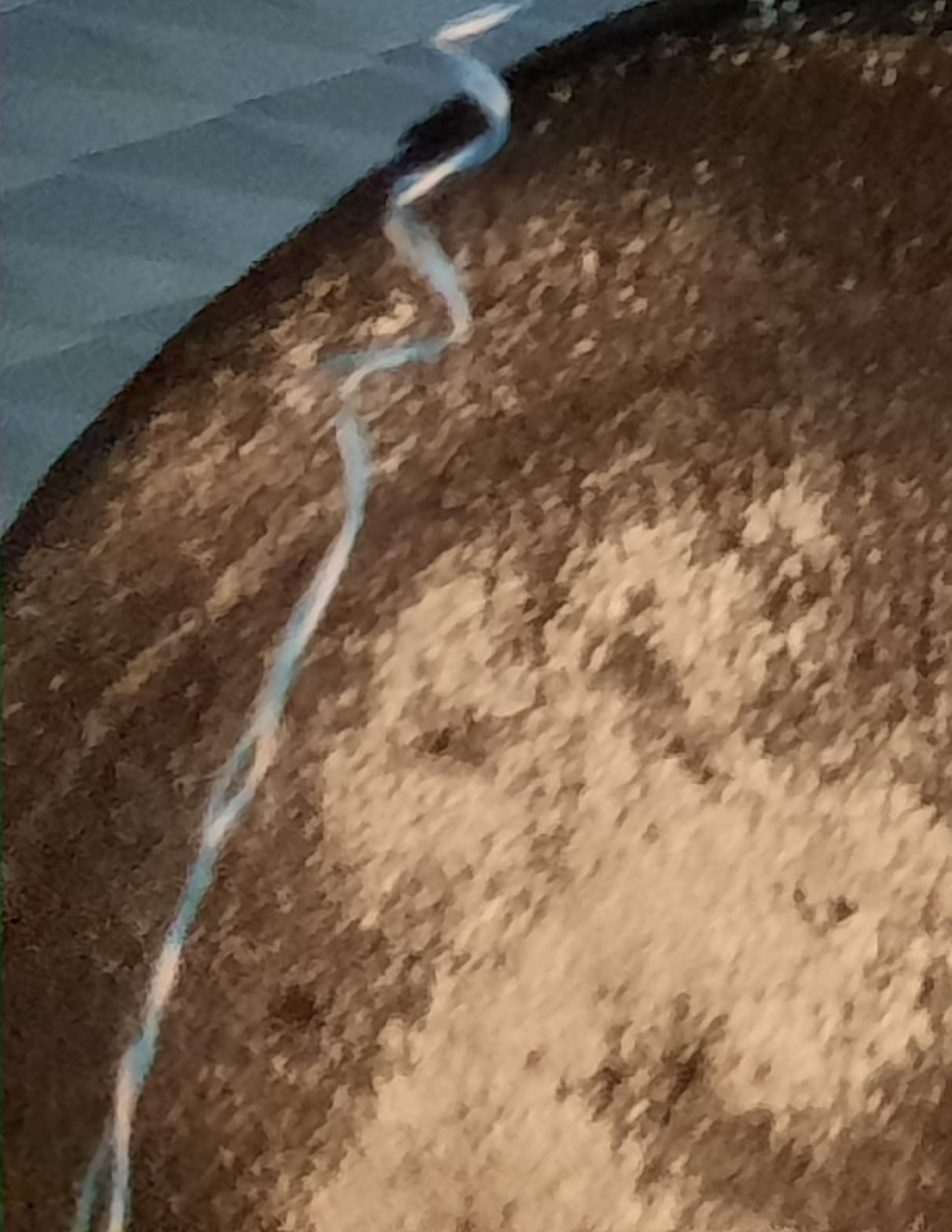}}\\[-2mm]
	\caption{(a) PMS PCB with the bias cables and the support; (b) Complete  measurement setup.}
	\label{fig:setup}
\end{figure}

\begin{figure}[h!]
	\centering
	\subfigure[]{\label{fig:rad_pattern1}\includegraphics[width=0.32\textwidth]{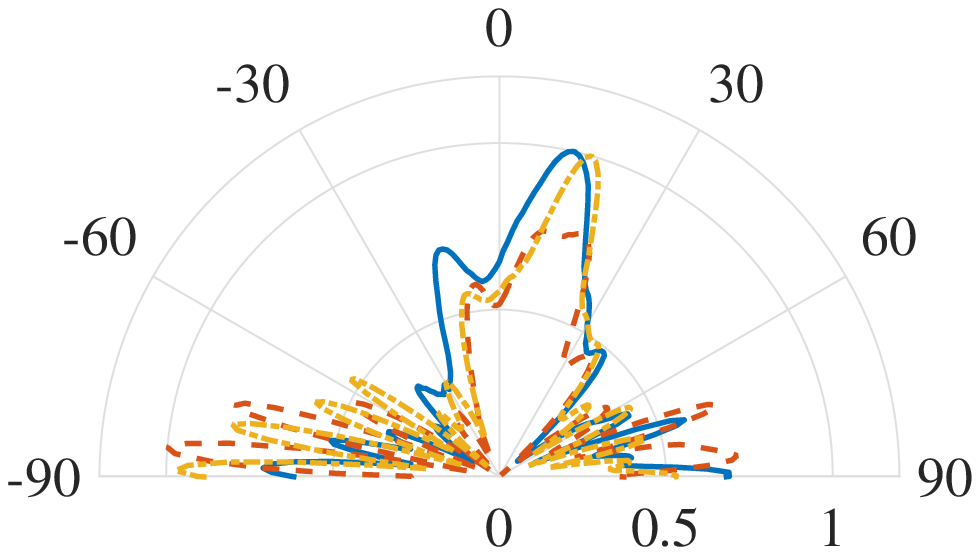}}
	\qquad
	\subfigure[]{\label{fig:rad_pattern2}\includegraphics[width=0.43\textwidth]{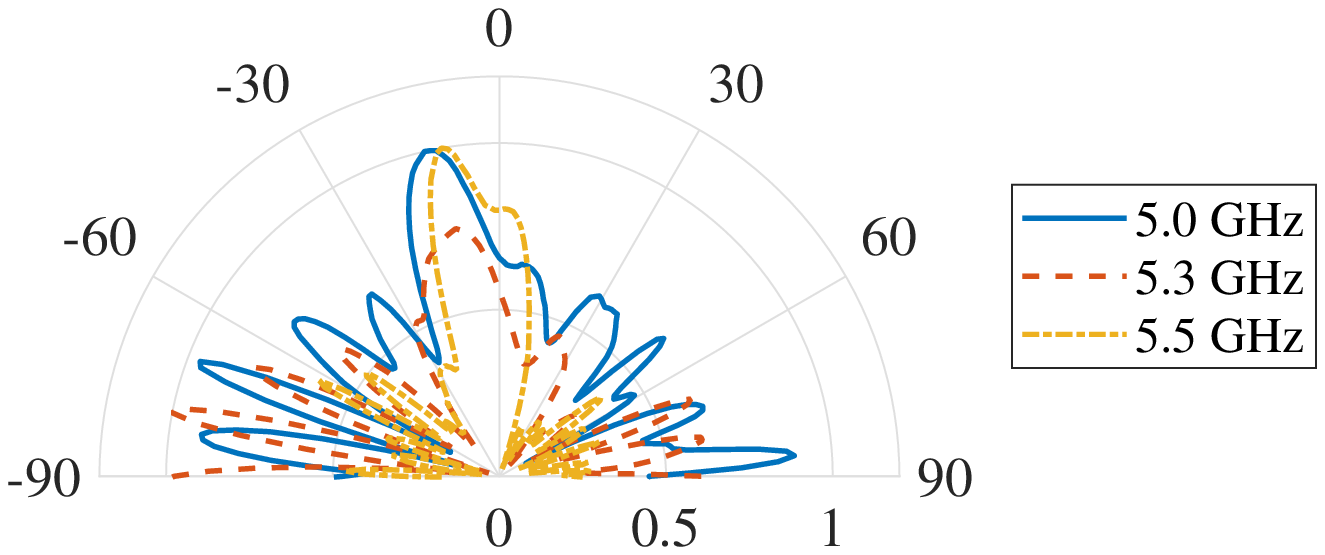}}\\[-3mm]
	\caption{Measured radiation patterns of the 3-by-10 PMS: (a) before rotation; (b) after a 180º rotation. }
	\label{fig:rad_pattern_meas}
\end{figure}

We have tested a few beamforming cases experimentally with a PMS prototype utilizing $3\times10$ unit cells. The PMS structure is illuminated by a horn antenna and the far-field pattern is formed after the incident field is reflected by the MS. The manufactured PCB of such PMS is depicted in Fig.~\ref{fig:PCB1}, and the measurement setup is shown in Fig.~\ref{fig:setup1}. The analytical model of this PMS was developed previously \cite{ConfTele2021, MM2021, PIERS, EExPolytech2021}. Commercially available varactor diodes MA46H120 have been used that provide a suitable capacitance variation between 0.12 and 1.0~pF, with a reverse bias voltage range from 1 to 11~V.
In our measurement setup, the PMS is placed at a distance $d_1 = 0.5$~m from the source horn antenna, as shown in
Fig.~\ref{fig:setup1}. The receiving antenna is placed at a distance of $d_2 \approx 5$~m from the PMS. Fig.~\ref{fig:rad_pattern_meas} shows the measured radiation patterns. The case in which the main beam points at $\theta = 15^\circ$ is depicted in Fig.~\ref{fig:rad_pattern1} with the capacitance variation from 0.13~pF (11~V bias) to 0.23 pF (6.7~V bias) along the MS. In Fig.~\ref{fig:rad_pattern2}, the main beam points at $\theta = -15^\circ$ under the same control voltage, however, the MS has been rotated by 180$^\circ$ around its normal in order to validate the results. The noticeable variation in the sidelobe levels is due to the higher controlling wires influence in the second configuration. Relatively high level of sidelobes can be also attributed to the rather small size of the realized MS prototype.

\section{Conclusion} 

This work outlines the theory of operation, the simulations and measurements of a reconfigurable beamforming PMS of a Sievenpiper mushroom type. In this work, we have selected a design that is suitable for operating at the microwave frequencies, with a unit cell geometry that allows convenient application of bias voltages to the controlling elements --- varactor diodes. In the proposed design, we use an implementation in which the controlling voltage is applied to the unit cells through a network of disconnected $x$- and $y$-coordinate lines. In this way, the radiation pattern flexibility is achieved. The proposed design has been studied numerically and tested experimentally with a MS prototype. The optimal design parameters have been determined for 5G applications in the 3--6 GHz frequency range. The performance of the PMS has been tested in a bi-static configuration with two horn antennas operating as a transmitter and a receiver. The obtained results validate the developed analytical and numerical methods and confirm potential benefits of using the studied PMS structures for the beam steering and beamforming applications.

\acknowledgement This work has been funded by Funda\c{c}\~{a}o para a
Ci\^{e}ncia e a Tecnologia (FCT), Portugal, under the Carnegie Mellon
Portugal Program (project ref.~CMU/TIC/0080/2019).


\end{document}